\documentclass[fleqn,usenatbib,letters]{mnras}

\usepackage{newtxtext,newtxmath}

\usepackage[T1]{fontenc}

\usepackage{graphicx}	
\usepackage{amsmath}	
\usepackage{multirow}
\usepackage[table]{xcolor}

\newcommand{\nq}[1]{%
	\begin{tabular}{@{}c@{}}\strut#1\strut\end{tabular}%
}

\title[New cosmological bounds on hot relics: Axions $\&$ Neutrinos]{New cosmological bounds on hot relics: Axions $\&$ Neutrinos}

\author[W. Giarè, E. Di Valentino, A. Melchiorri and O. Mena]{
William Giarè,$^{1}$\thanks{E-mail: william.giare@uniroma1.it }
Eleonora Di Valentino,$^{2}$\thanks{E-mail: eleonora.di-valentino@durham.ac.uk}
Alessandro Melchiorri,$^{1}$\thanks{E-mail: alessandro.melchiorri@roma1.infn.it}
and Olga Mena,$^{3}$\thanks{E-mail: omena@ific.uv.es}
\\
$^{1}$Physics Department and INFN, Universit\`a di Roma ``La Sapienza'', Ple Aldo Moro 2, 00185, Rome, Italy\\
$^{2}$Institute for Particle Physics Phenomenology, Department of Physics, Durham University, Durham DH1 3LE, UK\\
$^{3}$ IFIC, Universidad de Valencia-CSIC, 46071, Valencia, Spain
}

\pubyear{2020}

\begin{document}
\label{firstpage}
\pagerange{\pageref{firstpage}--\pageref{lastpage}}
\maketitle

\begin{abstract}
	Axions, if realized in nature, can be copiously produced in the early universe via thermal processes, contributing to the mass-energy density of thermal hot relics. In light of the most recent cosmological observations, we analyze two different thermal processes within a realistic mixed hot-dark-matter scenario which includes also massive neutrinos. Considering the axion-gluon thermalization channel we derive our most constraining bounds on the hot relic masses $m_a < 7.46$ eV and $\sum m_\nu< 0.114$~eV both at 95\% CL; while studying the axion-pion scattering, without assuming any specific model for the axion-pion interactions and remaining in the range of validity of the chiral perturbation theory, our most constraining bounds are improved to $m_a<0.91$~eV and $\sum m_\nu< 0.105$~eV, both at 95\% CL. Interestingly, in both cases, the total neutrino mass lies very close to the inverted neutrino mass ordering prediction. If future terrestrial double beta decay and/or long baseline neutrino experiments find that the nature mass ordering is the inverted one, this could rule out a wide region in the currently allowed thermal axion window. Our results therefore strongly support multi-messenger searches of axions and neutrino properties, together with joint analyses of their expected sensitivities.
\end{abstract}



\section{Introduction} \label{sec.Introduction}

The most elegant solution at present to the strong CP problem in Quantum Chromodynamics (QCD) would require the Lagrangian of the Standard Model of elementary particles to be invariant under an additional global $U(1)_{\rm PQ}$ (Peccei-Quinn) symmetry, spontaneously broken at some energy scale $f_{\rm a}$~\citep{Weinberg:1975ui,Nanopoulos:1973wz,Weinberg:1973un,Belavin:1975fg,Callan:1976je,Jackiw:1976pf,Shifman:1979if,Kim:1979if,Dine:1981rt,Peccei:1988ci,Peccei:2006as,Peccei:1977hh,Peccei:1977ur,Wilczek:1977pj}, with an associated Pseudo Nambu Goldestone Boson (PNGB), the so-called axion~\citep{Weinberg:1977ma,Kim:1986ax,Shifman:1979if,Kim:1979if,Dine:1981rt,Cheng:1987gp,Peccei:1988ci,Peccei:2006as,Marsh:2015xka, DiLuzio:2020wdo,Sikivie:2006ni}. A strong experimental effort has been devoted in different fields to search for axions~\citep{PhysRevD.40.3153,PhysRevLett.69.2333,PhysRevLett.64.2988,PhysRevD.42.1297,Irastorza:2018dyq,Anastassopoulos:2017ftl,Shilon:2012te,Shilon:2013xma,McAllister:2017lkb,2020arXiv200310894B,Lawson:2019brd,Du:2018uak,Braine:2019fqb}. If these elusive particles exist, they can be copiously produced via both thermal and non-thermal processes in the early universe. Axions produced via non-thermal processes, \emph{e.g.} by the vacuum realignment mechanism \citep{Abbott:1982af,Dine:1982ah,Preskill:1982cy,Linde:1985yf,Seckel:1985tj,Lyth:1989pb,Linde:1990yj} and/or by topological defects decay~\citep{Vilenkin:2000jqa,Kibble:1976sj,Kibble:1982dd,Vilenkin:1981kz,Davis:1986xc,Vilenkin:1982ks, Sikivie:1982qv,Vilenkin:1982ks,Huang:1985tt}, are natural cold dark matter candidates. Conversely, thermal axions, \emph{i.e.} the population of axions created and annihilated during interactions among particles in the primordial universe, contribute to the hot dark matter component instead. 

We shall focus here on the thermal axion scenario~\citep{Melchiorri:2007cd,Hannestad:2007dd,Hannestad:2008js,Hannestad:2010yi,Archidiacono:2013cha,Giusarma:2014zza,DiValentino:2015zta,DiValentino:2015wba,Archidiacono:2015mda,Hannestad:2005df}. 
While still relativistic, thermal axions, as other hot relics, behave as extra dark radiation, contributing to the effective number of relativistic degrees of freedom $N_{\rm eff}$,  defined by the relation
\begin{equation}
\rho_{\mathrm{rad}}=\left[1+\frac{7}{8}\left(\frac{4}{11}\right)^{4 / 3} N_{\mathrm{eff}}\right] \rho_{\gamma}~,
\end{equation}
with $\rho_{\gamma}$ the present Cosmic Microwave Background (CMB) energy-density. The reference value is $N_{\rm eff}=3.045$~\citep{Mangano:2005cc,2016JCAP...07..051D,Akita:2020szl,Froustey:2020mcq,Bennett:2020zkv}, and a departure from this standard scenario
may leave different signatures in the cosmological observables, modifying the damping tail of the CMB temperature angular power spectrum and changing two important scales at recombination: the sound horizon and the Silk damping scales. In addition, the primordial abundances of light elements predicted by the Big Bang Nucleosynthesis (BBN) are also sensitive to extra light species, \emph{i.e.} to a larger value of $N_{\rm eff}$. Indeed the expansion rate of the universe during the BBN epoch strongly depends on the effective number of relativistic degrees of freedom and extra light species, which will increase the expansion rate and lead to a higher freeze-out temperature for weak interactions, implying a higher fraction of primordial helium.  

When thermal axions become non-relativistic particles they leave identical signatures in the different cosmological observables as massive neutrinos, increasing the amount of the (hot) dark matter mass-energy density in our universe, suppressing structure formation at scales smaller than their free-streaming scale and leaving an imprint on the CMB temperature anisotropies, via the early integrated Sachs-Wolfe effect. This is why a large degeneracy between the axion and the total neutrino masses is expected~\citep{ Giusarma:2014zza,DiValentino:2015wba}. 

The final release of Planck 2018 temperature and polarization data~\citep{Akrami:2018vks}, offers a unique opportunity to derive updated bounds on the thermal axion mass, accounting also for the fact that neutrinos are massive particles, as robustly indicated by oscillation experiments~\citep{2020arXiv200611237D,deSalas:2018bym}: cosmology provides the most powerful mean to constrain their masses~\citep{deSalas:2018bym,2020arXiv200302289H,2019arXiv190708010V,Vagnozzi:2018pwo,Vagnozzi:2018jhn,Vagnozzi:2017ovm,Giusarma:2016phn,Bond:1980ha}. The paper is organized as follows. In Sec.~\ref{sec.Theory} we review the thermal processes of interest for this work; in Sec.~\ref{sec.Method} we describe our analysis method; in Sec.~\ref{sec.Results} we discuss our results; finally, in Sec.~\ref{sec.Conclusions} we present our conclusions. 

\section{Thermal Axions} \label{sec.Theory}
The thermal axion scenario can be described by two parameters: the axion coupling constant $f_a$ and the axion mass $m_a$, related as
\begin{equation}
m_{a}=\frac{f_{\pi} m_{\pi}}{f_{a}} \frac{\sqrt{R}}{1+R}\simeq 0.6\, \mathrm{eV} \times \frac{10^{7}
	\,\mathrm{GeV}}{f_{a}}~,
\end{equation}
where $R\doteq m_u/m_d\simeq0.553 \pm 0.043 $ is the up-to-down quark mass ratio and $f_{\pi}\simeq 93$ MeV is the pion decay constant \citep{Zyla:2020zbs}.  Axions decouple from the thermal bath when the reaction rate $\Gamma_a$ falls below the Hubble expansion rate 
\begin{equation}
H(T)=\sqrt{\frac{4 \pi^{3}}{45} g_{\star}(T)} \left(\frac{T^{2}}{M_{p l}}\right).
\end{equation}
with $M_{pl}\simeq 1.22 \times 10^{19} \rm{GeV}$ the Planck Mass and $g_{\star}(T)$ the relativistic degrees of freedom. Considering only the two-body processes with cross sections $\sigma_i = \sigma(p_i a \leftrightarrow p_jp_k)$ and with all the particles in thermal equilibrium, we can define the decay rate as \citep{DiLuzio:2020wdo}
\begin{equation}
\Gamma_{a}=\sum_{i} n_{i}\left\langle v \sigma_{i}\right\rangle
\end{equation}
where $n_i$ is the number density of $p_i$, $v\simeq 1$ the relativistic velocity and the brackets denote a thermal average. Solving the usual freeze-out condition 
\begin{equation}
\Gamma_{a}(T_d)=H(T_d)~,
\label{Freezout}
\end{equation}
one can estimate the decoupling temperature of a thermal axion population with mass $m_a$, while the axion contribution to the relativistic degrees of freedom is simply given by $g_{\star}(T_d)=g_{\star}^{\rm SM}(T_d)+1$. After decoupling ($T<T_d$)  axions maintain a thermal distribution which basically remains unaffected by other phenomena occurring in the plasma. Therefore we can estimate the current axion number density simply as
\begin{equation}
n_{a}=\frac{g_{\star S}\left(T_{0}\right)}{g_{\star S}\left(T_{d}\right)} \times \frac{n_{\gamma}}{2}~,
\label{na}
\end{equation}
with $n_{\gamma}\simeq 411 \mathrm{~cm}^{-3}$ the present photon density and $g_{\star S}(T_0)\simeq 3.91$ the current number of entropic degrees of freedom\footnote{We recall that  before the neutrino freeze-out $g_{\star S}=g_{\star}$}.

In the early Universe, there are several different processes that can produce distinct population of thermal axions. In this work we shall study a thermal axion population from axion-gluon and axion-pion scatterings separately.
\subsection{Axion-Gluon coupling}
In any QCD axion model, axions couple with free gluons. The relevant processes for axion thermalisation are 
\begin{itemize}
\item $a + q \leftrightarrow g  + q$ and  $a  + \bar{q} \leftrightarrow g + \bar{q}$;
\item $a + g \leftrightarrow q + \bar{q}$;
\item $a + g \leftrightarrow g + g$. 
\end{itemize}
Following \citet{DiLuzio:2020wdo}, the decoupling temperature for this population of thermal axions reads as 
\begin{equation}
T_{d} \simeq 4.0 \cdot 10^{11}\left(\frac{f_{a}}{10^{12} \mathrm{GeV}}\right)^{2} \mathrm{GeV}.
\end{equation}
It is easy to see that for $T_d \gg T_c$, we expect a contribution $n_a\simeq 7.5~\rm{cm}^{-3}$. It is also worth noting that the decoupling temperature must be smaller than the PQ breaking scale (above which there is no axion) and that if the scale of inflation and the reheating temperature are below $T_d$, this thermal population gets inflated away. In this paper, taking into account these caveats and exploiting current cosmological datasets, we constrain the sub-eV axion mass range allowed for this process in realistic scenarios that include also massive neutrinos.

\subsection{Axion-Pion Coupling}
After the QCD phase transition, $T<T_c$, axions can couple with hadrons. In practice, however, nucleons are so rare in the early universe with respect to pions that the only relevant process is the axion-pion interaction $\pi+\pi \leftrightarrow \pi + a $.

The leading order Lagrangian for axion-pion interaction reads
\begin{equation}
\mathcal{L}_{a \pi}=C_{a \pi} \frac{\partial_{\mu} a}{f_{a} f_{\pi}}\left(\pi^{0} \pi^{+} \partial_{\mu} \pi^{-}+\pi^{0} \pi^{-} \partial_{\mu} \pi^{+}-2 \pi^{+} \pi^{-} \partial_{\mu}~, \pi^{0}\right)
\label{Lap}
\end{equation}
where the axion-pion coupling 
\begin{equation}
C_{a \pi}=\frac{1}{3}\left(\frac{m_{d}-m_{u}}{m_{u}+m_{d}}+c_{d}^{0}-c_{u}^{0}\right)~,
\end{equation}
is a model-dependent quantity sensitive to the nature of axion-fermion interactions via the axion-quark couplings $c_{d}^{0}$ and $c_{u}^{0}$.
Starting from Eq.\eqref{Lap}, the leading order axion-pion interaction rate can be computed to obtain \citep{DiLuzio:2020wdo, 2021arXiv210110330D}
\begin{equation}
\Gamma_{a \pi}^{\rm LO} \simeq 0.215 C_{a \pi}^{2} \frac{T^{5}}{f_{a}^{2} f_{\pi}^{2}} h_{\rm{LO}}\left(\frac{m_{\pi}}{T}\right),
\label{Gamma_api}
\end{equation}
with $h(x)$ a rapidly decreasing function of its arguments normalized to $h(0)=1$. As usual, solving the freeze out condition \eqref{Freezout} we can estimate the decoupling temperature for an axion population with mass $m_a$, while by Eq.~\eqref{na} we can derive its current number density. 

However it should be noted that the thermal production of axions via pion scattering is strongly model-dependent since the relation between the axion mass and the (decoupling) temperature changes accordingly to the axion-pion interaction strength.  Consequently, the thermal production of axions from pion scattering could range between relatively large thermal abundances to negligible ones, depending on the precise value of $C_{a \pi}$. For example, in the KSVZ axion model \citep{Kim:1979if,Shifman:1979if,DiLuzio:2020wdo} the coupling between axions and fermions vanishes at tree level: $c_{d}^{0}=c_{u}^{0}=0$ and $C_{a\pi}=(1-R)/(3+3R)\simeq 0.12$ leading to a sizable amount of relic axions. On the other hand, in the DFSZ scenario \citep{Dine:1981rt,Zhitnitsky:1980tq,DiLuzio:2020wdo} because of the presence of extra Higgs doublets, QCD axions couple to SM fermions at tree level - $c_{u}^{0}=\frac{1}{3} \cos ^{2}(\beta)$, $c_{d}^{0}=\frac{1}{3} \sin ^{2}(\beta)$ with $\tan \beta \in[0.25,170]$ because of the unitary of tree-lever fermion scatterings \citep{DiLuzio:2020wdo} - and the axion production can be either enhanced or suppressed: $C_{a\pi}=(1-R)/(3+3R)-1/9\cos(2\beta)$, see the recent discussion by \citet{2020arXiv201206566F}.
 
Furthermore, the authors of Ref.~\citep{2021arXiv210110330D} have recently shown that for temperatures $T_{\chi}\gtrsim 62$ MeV the next-to-leading order term in the axion-pion interaction rate
 \begin{equation}
 \Gamma_{a \pi}^{\rm NLO} \simeq -0.62 C_{a \pi}^{2} \frac{T^{7}}{f_{a}^{2} f_{\pi}^{4}} h_{\rm{NLO}}\left(\frac{m_{\pi}}{T}\right),    
 \end{equation}
becomes comparable with the leading order part, $\Gamma_{a \pi}^{\rm NLO}(T_{\chi})\simeq 0.5 \, \Gamma_{a \pi}^{\rm LO}(T_{\chi})$, and the chiral perturbation theory breaks down. Interestingly, for the KSVZ model these controversial values for the temperatures precisely correspond to the sub-eV axion mass range of interest for current and future CMB experiments.  Therefore, it is mandatory to adopt a model-independent approach to be able to compute reliable thermal axion mass limits from cosmology until a robust lattice QCD method provides the precise answer for a given model in these temperature ranges. 

In order to study  the thermalisation from axion-pion scatterings in the most broad and reliable scenario, we restrict ourselves to explore exclusively to the parameter space where the next-to-leading order term $\Gamma_{a \pi}^{\rm NLO}(T)$ remains small with respect to leading order part $\Gamma_{a \pi}^{\rm LO}(T)$, which basically means to consider decoupling temperatures $T_d\lesssim T_{\chi}\simeq 62 \rm{MeV}$. 
In addition, we shall not assume in the following any specific underlying theoretical model for the axion-pion interactions, leaving the axion-pion coupling $C_{a\,\pi}$ as a free parameter. In this way, we are not only able to explore different axion models beyond the usual KSVZ and  DSFZ scenarios~\footnote{While the KSVZ and DFSZ are widely considered as benchmark scenarios, there are other models in which both the new heavy quarks and the Higgs doublets carry $U(1)_{\rm PQ}$ charges, see \textit{e.g.}~\citet{Kim:2008hd} and \citet{DiLuzio:2020wdo}.}, but also derive well-defined constraints on the sub-eV axion mass range in realistic scenarios which include also massive neutrinos.

\section{Numerical Analyses} \label{sec.Method}

\begin{table}
	\begin{center}
		\renewcommand{\arraystretch}{1.3}
		\begin{tabular}{c@{\hspace{0. cm}}@{\hspace{0.7 cm}} c @{\hspace{0.7 cm}} c }
			\hline
			\textbf{Parameter}                    & \textbf{Prior for axion-gluon}  & \textbf{Prior for axion-pion}\\
			\hline\hline
			$\Omega_{\rm b} h^2$         & $[0.005\,,\,0.1]$ &$[0.005\,,\,0.1]$\\
			$\Omega_{\rm c} h^2$       & $[0.005\,,\,0.1]$ &$[0.005\,,\,0.1]$\\
			$100\,\theta_{\rm {MC}}$             & $[0.5\,,\,10]$ &$[0.5\,,\,10]$\\
			$\tau$                       & $[0.01\,,\,0.8]$ & $[0.01\,,\,0.8]$\\
			$\log(10^{10}A_{\rm S})$         & $[1.61\,,\,3.91]$ & $[1.61\,,\,3.91]$\\
			$n_{\rm S}$                        & $[0.8\,,\, 1.2]$ & $[0.8\,,\, 1.2]$\\
			$\sum m_{\nu}\,[\rm{eV}]$ & $[0.06 \,,\,5]$ & $[0.06 \,,\,5]$\\ 
			$m_a\,[\rm{eV}]$ &$[0.1\,,\,10]$ &-\\
			$T_d\,[\rm{MeV}]$& - & $<62$\\
			$C_{a\pi}$ & -& $[0\,,\,0.5]$\\
			
			\hline\hline
		\end{tabular}
		\caption{List of the parameter priors.}
		\label{tab.Priors}
	\end{center}
	
\end{table}

The scenario we analyze is an extension of the $\Lambda$CDM model which includes both axions and neutrinos as hot thermal massive relics.  We perform Monte Carlo Markov Chain (MCMC) analyses using a modified version of the publicly available package \texttt{CosmoMC}~\citep{Lewis:2002ah,Lewis:2013hha} and computing the theoretical model with the latest version of the Boltzmann code \texttt{CAMB}~\citep{Lewis:1999bs,Howlett:2012mh}.  We consider the canonical $\Lambda$CDM model described by the usual six-parameters, \emph{i.e.}, the baryon $\omega_{\rm b}\equiv \Omega_{\rm b}h^2$ and cold dark matter $\omega_{\rm c}\equiv\Omega_{\rm c}h^2$ energy densities, the angular size of the horizon at the last scattering surface $\theta_{\rm{MC}}$, the optical depth $\tau$, the amplitude of primordial scalar perturbation $\log(10^{10}A_{\rm S})$ and the scalar spectral index $n_{\rm S}$. 
Together with the standard $\Lambda$CDM parameters, we add the thermal axion mass $m_a$ and the sum of three active neutrino masses $\sum m_{\nu}$ (both in $\rm{eV}$). For the axion-pion thermalization channel we consider also the coupling $C_{a\pi}$ and we restrict our scan only to decoupling temperatures $T_{d}\lesssim 62\,\rm{MeV}$ where $\Gamma^{\rm NLO}_{a\pi}(T)$ remains small with respect to $\Gamma^{\rm LO}_{a\pi}(T)$. In this case, for each sampled point $(T_d\,,\,C_{a\pi})$ we compute the axion mass $m_a(T_d\,,C_{a\pi})$ by solving Eq.\eqref{Freezout}. We vary these parameters in a range of external and conservative priors listed in Tab.~\ref{tab.Priors}.

The posteriors of our parameter space have been explored using the MCMC sampler developed for \texttt{CosmoMC} and tailored for parameter spaces with a speed hierarchy which also implements the "fast dragging" procedure described in Ref.~\citep{Neal:2005}. The convergence of the chains obtained with this procedure is tested using the Gelman-Rubin criterion~\citep{Gelman:1992zz} and we choose as a threshold for chain convergence $R-1 \lesssim 0.02 $. Our baseline data-set consists of:
\begin{itemize}
	
	\item Planck 2018 temperature and polarization (TT TE EE) likelihood, which also includes low multipole data ($\ell < 30$)~\citep{Aghanim:2019ame,Aghanim:2018eyx,Akrami:2018vks}. We refer to this combination as "Planck".
	
	\item Planck 2018 lensing likelihood~\citep{Aghanim:2018oex}, constructed from measurements of the power spectrum of the lensing potential. We refer to this dataset as "lensing".
	
	\item Baryon Acoustic Oscillations (BAO) measurements extracted from data from the 6dFGS~\citep{2011MNRAS.416.3017B}, SDSS MGS~\citep{Ross:2014qpa} and BOSS DR12~\citep{Alam:2016hwk} surveys. We refer to this dataset as "BAO".
	
	\item Type Ia Supernovae (SNeIa) distance moduli measurements from the Pantheon sample \citep{2018ApJ...859..101S}. We refer to this dataset as "Pantheon".
	
	\item Galaxy clustering and cosmic shear measurements, as well as their cross-correlations, from the Dark Energy Survey~\citep{Troxel:2017xyo,Abbott:2017wau,2017arXiv170609359K}. We refer to this dataset as "DES".
	
	\item The Hubble constant measurement from the SH0ES collaboration analysing type-Ia supernovae data from the Hubble Space Telescope~\citep{Riess:2020fzl}. We refer to this dataset as "R20".
	
\end{itemize}

\section{Results} \label{sec.Results}
In this section we present the results obtained by our MCMC analysis of the mixed hot dark matter scenario which includes axions and neutrinos as hot thermal massive relics. We consider both the axion-gluon and the axion-pion thermalization channels.

\begin{table*}
	\begin{center}
		\renewcommand{\arraystretch}{1.5}
		\resizebox{\textwidth}{!}{\begin{tabular}{ c|c|c|c|c|c|c|c|c }
				\hline
				\multirow{2}*{\textbf{DATASET}} &  \multicolumn{8}{c}{\textbf{AXION-GLUON COUPLING}} \\ \cline{2-9} & 
				\boldmath{$\Omega_{\rm b} h^2$} & 
				\boldmath{$\Omega_{\rm c} h^2$} & 
				\boldmath{$100\,\theta_{\rm {MC}}$} &
				\boldmath{$\tau$ } & 
				\boldmath{$\log(10^{10}A_{\rm S})$} & 
				\boldmath{$n_{\rm S}$}&\boldmath{$m_a\, [\rm eV]$}&
				\boldmath{$\sum m_{\nu}\, [\rm eV]$}\\
				\hline\hline 
				
				\textbf{Planck} & $0.02236\pm0.00016$ &$0.1188^{+0.0033}_{-0.0014}$ & $1.0476 \pm 0.00032$&$0.0546^{+0.0072}_{-0.0082}$&$3.047^{+0.015}_{-0.017}$&$0.9631\pm0.0046$&$< 8.35$&$< 0.324$\\
				\hline  
				\nq{\textbf{Planck}\\ \textbf{+lensing}} & $0.02236\pm0.00015$ &$0.1191^{+0.0030}_{-0.0012}$&$1.04076 \pm 0.00031$&$0.0553\pm0.0075$&$3.049 \pm 0.015$&$0.9626\pm0.0044$&$< 8.03$&$< 0.272$\\
				\hline
				\nq{\textbf{Planck}\\\textbf{+BAO}} & $0.02248\pm0.00014$ &$0.1176^{+0.0029}_{-0.00083}$&$1.04099 \pm 0.00029$&$0.0568^{+0.0072}_{-0.0083}$&$3.048^{+0.015}_{-0.017}$&$0.9672\pm0.0040$&$< 8.14$&$< 0.158$\\
				\hline  
				\nq{\textbf{Planck}\\\textbf{+Pantheon}}& $0.02242\pm0.00014$ &$0.1181^{+0.0034}_{-0.0011}$&$1.04086 \pm 0.00031$&$0.0554^{+0.0073}_{-0.0085}$&$3.046^{+0.015}_{-0.017}$&$0.9648\pm0.0044$&$< 8.62$&$< 0.209$\\
				\hline  
				\nq{\textbf{Planck}\\\textbf{+DES}} & $0.02248\pm0.00015$ &$0.1160^{+0.0028}_{-0.0016}$&$1.04093 \pm 0.00032$&$0.0549\pm0.0079$&$3.043 \pm 0.016$&$0.9661\pm 0.0044$&$< 8.40$&$<0.346$\\
				\hline  
				\nq{\textbf{Planck}\\\textbf{+R20}}& $0.02258\pm0.00015$ &$0.1168^{+0.0028}_{-0.00097}$&$1.04113 \pm 0.00030$&$0.0579^{+0.0073}_{-0.0082}$&$3.048\pm 0.017$&$0.9697\pm0.0043$&$<7.92$&$< 0.129$\\
				\hline  
				\nq{\textbf{Planck} \textbf{+lensing}\\\textbf{+BAO} \textbf{+DES} \\ \textbf{+Pantheon} } & $0.02255\pm0.00013$ &$0.1159^{+0.0029}_{-0.0012}$&$1.04105 \pm 0.00029$&$0.0594^{+0.0068}_{-0.0079}$&$3.052^{+0.013}_{-0.016}$&$0.9677\pm0.0038$&$< 8.13$&$< 0.136$\\
				\hline  
				\nq{\textbf{Planck}  \textbf{ +lensing}\\\textbf{+BAO} \textbf{+DES} \\\textbf{+Pantheon} \textbf{+R20}} & $0.02265\pm0.00013$ &$0.1156^{+0.0026}_{-0.0010}$&$1.04118 \pm 0.00030$&$0.0624^{+0.0073}_{-0.0087}$&$3.057^{+0.015}_{-0.017}$&$0.9701\pm0.0038$&$< 7.46$&$< 0.114$\\
				\hline\hline
		\end{tabular}}
		\caption{Results for the Axion-Gluon thermalization channel obtained for different combination of the datasets listed in Sec.~\ref{sec.Method}. The bounds on parameters are 1$\sigma$ errors ($68\%$ CL), while the upper bounds are 2$\sigma$ ($95\%$ CL) constraints.}
		
		\label{tab.Results1}
	\end{center}
\end{table*}

\begin{table*}
	\begin{center}
		\renewcommand{\arraystretch}{1.5}
		\resizebox{\textwidth}{!}{\begin{tabular}{ c|c|c|c|c|c|c|c|c }
				\hline
				\multirow{2}*{\textbf{DATASET}} &  \multicolumn{8}{c}{\textbf{AXION-PION COUPLING,   \boldmath{$T_d<62\,\rm{MeV}$}}} \\ \cline{2-9} & 
				\boldmath{$\Omega_{\rm b} h^2$} & 
				\boldmath{$\Omega_{\rm c} h^2$} & 
				\boldmath{$100\,\theta_{\rm {MC}}$} &
				\boldmath{$\tau$ } & 
				\boldmath{$\log(10^{10}A_{\rm S})$} & 
				\boldmath{$n_{\rm S}$}&\boldmath{$m_a\, [\rm eV]$}&
				\boldmath{$\sum m_{\nu}\, [\rm eV]$}\\
				\hline\hline 
				
				\textbf{Planck} & $0.02261\pm0.00015$ &$0.1252^{+0.0022}_{-0.0016}$ & $1.04012 \pm 0.00032$&$0.0567^{+0.0072}_{-0.0083}$&$3.064^{+0.015}_{-0.017}$&$0.9703\pm0.0058$&$< 2.41$&$< 0.269$\\
				\hline  
				\nq{\textbf{Planck}\\ \textbf{+lensing}} & $0.02257\pm0.00015$ &$0.1266^{+0.0016}_{-0.0014}$&$1.04004 \pm 0.00032$&$0.0592^{+0.0072}_{-0.0087}$&$3.071^{+0.014}_{-0.017}$&$0.9706^{+0.0056}_{-0.0049}$&$< 1.96$&$< 0.221$\\
				\hline
				\nq{\textbf{Planck}\\\textbf{+BAO}} & $0.02279\pm0.00014$ &$0.1238\pm0.0012$&$1.04043 \pm 0.00029$&$0.0609^{+0.0075}_{-0.0091}$&$3.066^{+0.015}_{-0.018}$&$0.9819\pm0.0040$&$< 1.04$&$< 0.134$\\
				\hline  
				\nq{\textbf{Planck}\\\textbf{+Pantheon}}& $0.02268\pm0.00015$ &$0.1250^{+0.0017}_{-0.0015}$&$1.04023 \pm 0.00031$&$0.0582^{+0.0072}_{-0.0086}$&$3.064^{+0.015}_{-0.017}$&$0.9758^{+0.0052}_{-0.0046}$&$< 1.78$&$< 0.169$\\
				\hline  
				\nq{\textbf{Planck}\\\textbf{+DES}} & $0.02270\pm0.00014$ &$0.1239\pm0.0013$&$1.04024 \pm 0.00031$&$0.0568^{+0.0076}_{-0.0085}$&$3.061^{+0.015}_{-0.017}$&$0.9727\pm 0.0056$&$<2.16$&$<0.257$\\
				\hline  
				\nq{\textbf{Planck}\\\textbf{+R20}}& $0.02281\pm0.00014$ &$0.1240\pm0.0013$&$1.04041 \pm 0.00030$&$0.0608^{+0.0077}_{-0.0092}$&$3.067^{+0.016}_{-0.018}$&$0.9811\pm0.0043$&$<1.17$&$< 0.124$\\
				\hline  
				\nq{\textbf{Planck} \textbf{+lensing}\\\textbf{+BAO} \textbf{+DES} \\ \textbf{+Pantheon} } & $0.02286\pm0.00013$ &$0.1233^{+0.0010}_{-0.00090}$&$1.04047 \pm 0.00029$&$0.0683^{+ 0.0081}_{-0.0095}$&$3.082^{+0.016}_{-0.018}$&$0.9828\pm0.0037$&$< 1.04$&$< 0.115$\\
				\hline  
				\nq{\textbf{Planck}  \textbf{ +lensing}\\\textbf{+BAO} \textbf{+DES} \\\textbf{+Pantheon} \textbf{+R20}}  & $0.02292\pm0.00013$ &$0.12271\pm0.00091$&$1.04059 \pm 0.00028$&$0.0706^{+ 0.0084}_{-0.010}$&$3.085^{+0.016}_{-0.019}$&$0.9848\pm0.0036$&$< 0.91$&$< 0.105$\\
				\hline\hline
		\end{tabular}}
		\caption{Results for the Axion-Pion thermalization channel obtained for different combination of the datasets listed in Sec.~\ref{sec.Method}. The bounds on parameters are 1$\sigma$ errors ($68\%$ CL), while the upper bounds are 2$\sigma$ ($95\%$ CL) constraints.}
		
		\label{tab.Results2}
	\end{center}
\end{table*}	

\begin{figure*}
	\centering
	\includegraphics[width=0.95 \textwidth]{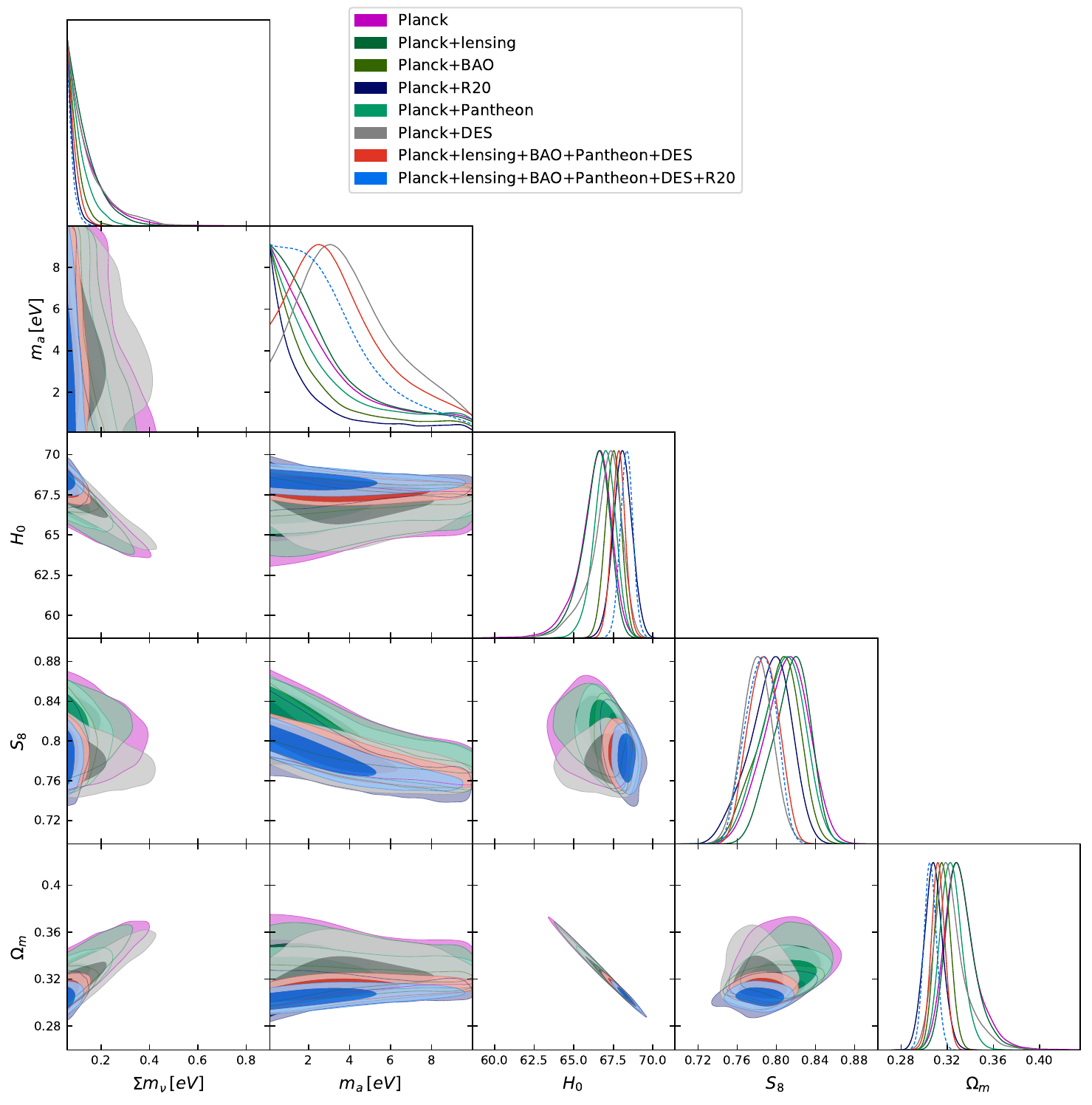}
	\caption{Axion-Gluon thermalization channel. Marginalized 2D and 1D posteriors for different combinations of the datasets listed in Sec.~\ref{sec.Method}.}
	\label{fig:figure1}
\end{figure*}

\begin{figure}
	\centering
	\includegraphics[width=0.45 \textwidth]{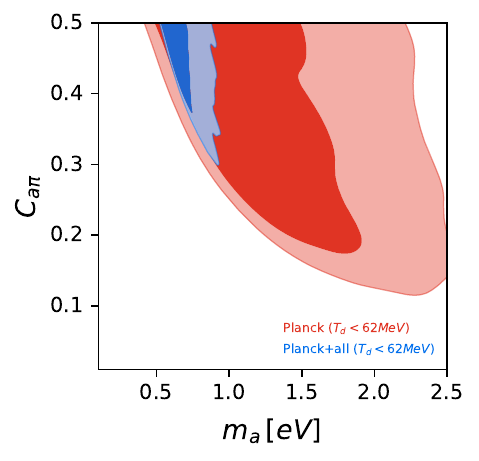}
	\caption{Axion-Pion thermalization channel. Marginalized 2D and 1D posteriors in the plane $(m_a\,,\,C_{a\pi})$ with the prior $T_d<62\,\rm{MeV}$.}
	\label{fig:figure2}
\end{figure}

\begin{figure*}
	\centering
	\includegraphics[width=0.95 \textwidth]{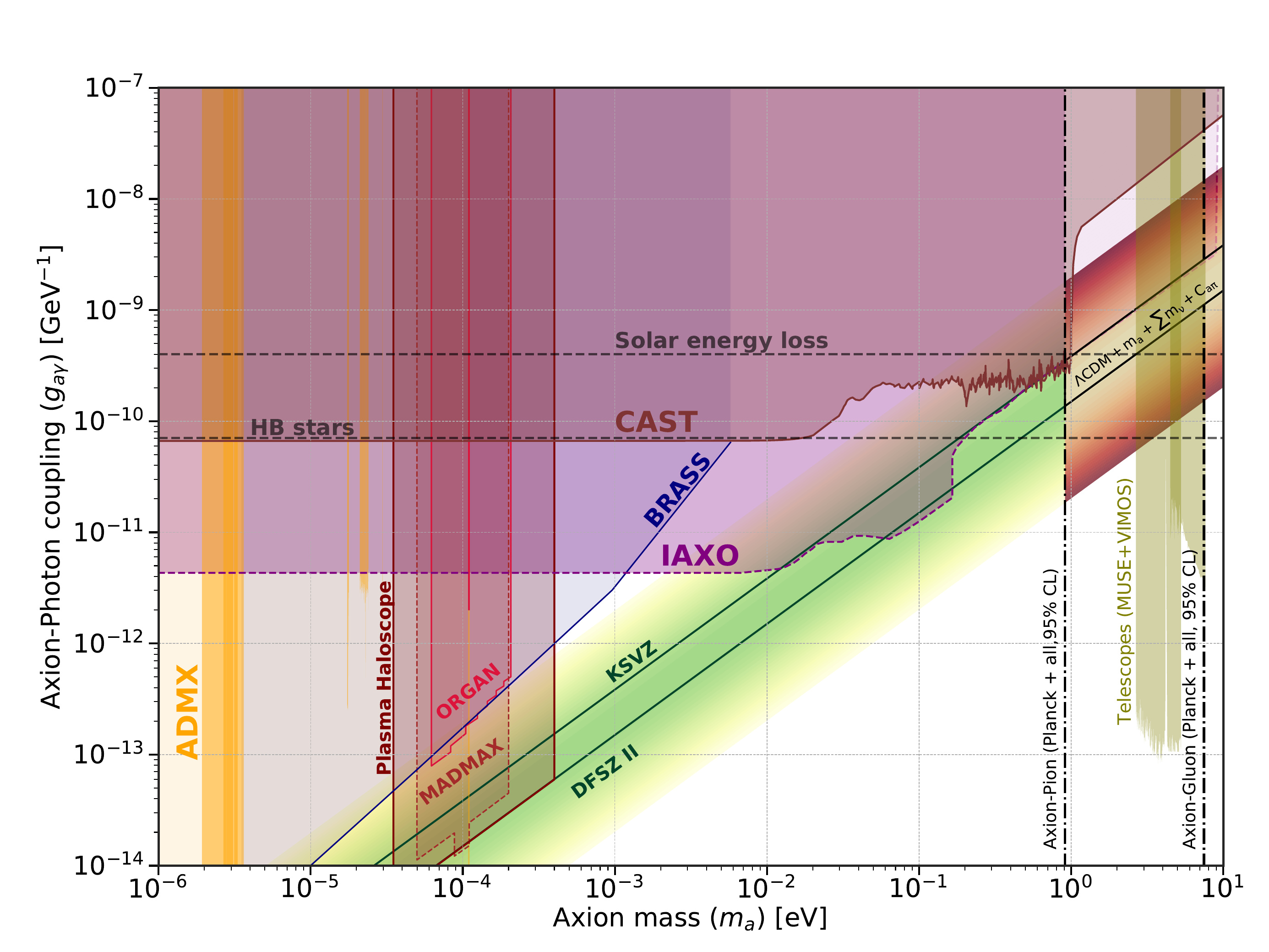}
	\caption{Axion limits in the plane ($m_a$ , $g_{a\gamma})$. We quote our most constraining cosmological bounds (both at 95\% CL) for the axion-pion and the axion-gluon thermalization channels. We also show current limits and future detection sensitivity forecasts for different experiments: CAST~\citep{Anastassopoulos:2017ftl}, IAXO~\citep{Shilon:2012te,Shilon:2013xma}, ORGAN~\citep{McAllister:2017lkb}, MADMAX~\citep{2020arXiv200310894B}, Plasma Haloscope~\citep{Lawson:2019brd}, ADMX~\citep{Du:2018uak,Braine:2019fqb}, Telescopes~\citep{Grin:2006aw,Regis:2020fhw} and  \href{http://wwwiexp.desy.de/groups/astroparticle/brass/brassweb.htm}{BRASS}. The horizontal dashed lines represent the limits from the Sun and horizontal branch (HB) stars energy loss \citep{Ayala:2014pea}.}
	\label{fig:figure3}
\end{figure*}

\subsection{Axion-Gluon scatterings}

Table~\ref{tab.Results1} summarizes the results for the axion-gluon thermalization channel obtained from our MCMC analyses of the $\Lambda\rm{CDM} + \sum m_{\nu} + m_a$ model. Figure~\ref{fig:figure1} shows the $68\%$ and $95\%$~CL contour plots for different cosmological parameters. 

As discussed in the introduction, a large degeneracy is expected between the axion and neutrino masses, see also Fig.~\ref{fig:figure1}, where a strong anticorrelation is clearly noticed from the allowed contours in the ($\sum m_\nu$, $m_a$) plane. Furthermore, these two parameters show similar degeneracies with other cosmological parameters and quantities such as $H_0$, $S_8$ and $\Omega_m$.

Exploiting the last release of Planck's temperature and polarization (TT,TE,EE+lowP) data, we derive the 95\% CL upper bounds $\sum m_{\nu}<0.324\,\rm{eV}$ and $m_a<8.35\,\rm{eV}$. Notice that due to the lower contribution in $\Omega_a\,h^2$ and $\Delta N_{\rm eff}$ expected by the axion-pion thermalization channel, the bounds on the axion mass are much less tight than those presented for the axion-pion case in Ref.~\citep{DiValentino:2015wba}. Indeed for the axion-gluon thermalization channel, eV axion masses correspond to high decoupling temperatures $T_{d}\gg T_c$ and for $T_d\gtrsim 150$ GeV all particles of the Standard Model are relativistic so that $g_{\star}\simeq 107,75$. From Eq. \eqref{na} one can notice that this will lead to a number density of relic axion $n_a\simeq7.5\,\rm{cm}^{-3}$ (that does not depend on the decoupling temperature), giving a very small contribution $\Delta N_{\rm{eff}}\simeq 0.027$ to the effective number of relativistic degrees of freedom, which is well beyond the constraining power of Planck data.

As concerns the other datasets involved in our analyses, we notice that the axion mass bounds only weakly change with the dataset. For example, the inclusion of CMB lensing measurements from the Planck satellite only slightly improves the neutrino mass bound to $\sum m_{\nu}<0.272\,\rm{eV}$ at 95\% CL, leaving the constraints on the axion mass almost unchanged ($m_a<8.03\,\rm{eV}$ at 95\% CL). Instead, the combination of Planck and DES data slightly worsens both the bounds on the axion mass ($m_a<8.40\,\rm{eV}$ at 95\% CL), and the  bounds on neutrinos ($\sum m_{\nu}<0.346 \,\rm{eV}$ at 95\% CL). This is due to the lower value of the clustering parameter $\sigma_8$ preferred by DES measurements, which is translated into a higher hot thermal relic masses. Conversely, due to the smaller error in $\Omega_m$, the inclusion of Pantheon data leads to a significant improvement in the constraints on the sum of neutrino masses ($\sum m_{\nu}<0.209\,\rm{eV}$ at 95\%CL), but not in those on the axion mass ( $m_a<8.62\,\rm{eV}$, at 95\% CL). 

As expected, the largest impact on Planck bounds arises from the inclusion of the large-scale structure information from BAO measurements. As also discussed by~\citet{DiValentino:2015wba}, hot thermal particles as axions and neutrinos suppress structure formation at small scales and therefore galaxy clustering information becomes crucial to set bounds on the amount of dark matter in the form of these relics. Indeed, combining Planck and BAO data we derive the upper bounds $m_a<8.14 \,\rm{eV}$ and $\sum m_{\nu}<0.158\,\rm{eV}$, both at 95\% CL. Combining together all the datasets, we obtain the robust 95\% CL upper limits of $m_a<8.13\,\rm{eV}$ and $\sum m_{\nu}<0.136\,\rm{eV}$.

Despite the fact that there is a very large tension among CMB estimates and low redshift measurements of the Hubble constant - with a statistical significance above $4\sigma$~\citep{Verde:2019ivm,2020arXiv200811284D,DiValentino:2021izs} -, this tension is considerably reduced in the presence of additional relativistic degrees of freedom. Sub-eV thermal axions are relativistic at decoupling and therefore will ease the well-known Hubble constant tension. Consequently, the addition of the prior on $H_0$ as measured by the Hubble Space Telescope in our cosmological analyses is perfectly justified and leads to a further strong improvement in the constraints on $\sum m_{\nu}$. The reason beyond this improvement can be easily understood in terms of the large degeneracy between the neutrino masses and the Hubble constant. It is well known that an increase on $\sum m_{\nu}$ induces a shift in the distance to last scattering that can be easily compensated by lowering $H_0$, leading to a strong degeneracy between these two parameters. Such a degeneracy  can be broken by an independent measurement of $H_0$ as that provided by the SH0ES Collaboration. Combining R20 and Planck data leads to upper bounds on the thermal relic masses of $\sum m_{\nu}<0.129\,\rm{eV}$ and $m_a<7.92\,\rm{eV}$, both at 95\% CL. Including also Planck lensing measurements, BAO, Pantheon and DES data, the upper bound on the neutrino mass becomes $\sum m_{\nu}<0.114\,\rm{eV}$ at 95\% CL, while the bound on the axion mass  is slightly improved to $m_a<7.46\,\rm{eV}$ at 95\% CL. Notice that the former upper limit on the total neutrino mass is very close to the inverted neutrino mass ordering prediction, implying that a future measurement  of the nature's mass ordering could be translated into a limit on the thermal axion parameter space. 

\subsection{Axion-Pion scatterings}

We shall now focus on the axion-pion thermalization channel. In this case the chiral perturbation theory adopted to compute the abundance of relic axions produced via pion scattering becomes unsafe for values of the decoupling temperatures above $T_{\chi}\simeq 62\,\rm{MeV}$ \citep{2021arXiv210110330D}. For any axion model, this limit defines the smallest mass which can be safely explored within a perturbative approach:
\begin{equation}
m_a\lesssim m_a(T_{\chi},C_{a\pi})\simeq 1.2 \times \left(\frac{0.12}{C_{a\pi}}\right)\,\rm{eV}.
\label{axion_mass_limit}
\end{equation} 
Notice that when $m_a\lesssim 1.2 \times\left(0.12/C_{a\pi}\right)\,\rm{eV} $ any bound derived using effective field theory is not completely reliable. Until robust lattice QCD methods provide a definitive answer, we have basically two choices: either we assume that when temperatures exceed $62\,\rm{MeV}$ perturbation theory still provides a reasonable approximation of a more accurate non-perturbative result, or, more conservatively, we limit our scan only to temperatures below $62\,\rm{MeV}$.  Here, we present and discuss the results obtained following the latter more conservative approach.

Table~\ref{tab.Results2} summarizes the constraints for the model $\Lambda\rm{CDM} + \sum m_{\nu} + m_a + C_{a\pi}$ for the different datasets listed in Sec.\ref{sec.Method}. Figure \ref{fig:figure2} clearly illustrates the fact that requiring $T_d<62 \,\rm{MeV}$ implies less constraining bounds on the axion mass. We estimate the upper bound on the axion mass as the value which corresponds to the 95\% of its integrated posterior distribution function. Therefore, we derive strong conservative bounds without extending the theory in a region beyond its validity.

Exploiting the last release of Planck temperature and polarization (TT,TE,EE+lowP) data, we derive the upper bound $m_a<2.41\,\rm{eV}$ at 95\% CL on the axion mass and $\sum m_{\nu}<0.269\,\rm{eV}$ at 95\% CL on neutrinos.

As concerns the other datasets considered in our analyses, in this case their impact on the axion-mass bounds is relevant. Indeed, we may appreciate that the inclusion of CMB lensing measurements from the Planck satellite improve both the neutrino mass bound ($\sum m_{\nu}<0.221\,\rm{eV}$ at 95\% CL) and the constraints on the axion mass  ($m_a<1.96\,\rm{eV}$ at 95\% CL). Due to the lower value of the clustering parameter $\sigma_8$ preferred by DES measurements, in this case the combination of Planck and DES data gives $m_a<2.16\,\rm{eV}$ and $\sum m_{\nu}<0.257 \,\rm{eV}$, both at 95\% CL. On the other hand, the smaller error in $\Omega_m$ of Pantheon data leads to an improvement both in the constraints on the sum of neutrino masses ($\sum m_{\nu}<0.169\,\rm{eV}$ at 95\%CL), and in the constraints on the axion mass ( $m_a<1.78\,\rm{eV}$, at 95\% CL). 

Once again, the largest impact on Planck bounds arises from the inclusion of large-scale structure information from BAO measurements. Indeed, combining Planck and BAO data we derive the upper bounds $m_a<1.04 \,\rm{eV}$ and $\sum m_{\nu}<0.134\,\rm{eV}$, both at 95\% CL. Combining together all the aforementioned datasets, we obtain the very tight and robust 95\% CL upper limits of $m_a<1.04\,\rm{eV}$ and $\sum m_{\nu}<0.115\,\rm{eV}$. 

Considering also the prior on $H_0$ as measured by the Hubble Space Telescope and combining together the R20 and Planck data, we obtain the upper bounds on the thermal relic masses $\sum m_{\nu}<0.124\,\rm{eV}$ and $m_a<1.17\,\rm{eV}$, both at 95\% CL. Including also Planck lensing measurements, BAO, Pantheon and DES data the upper bound on the neutrino mass is improved to $\sum m_{\nu}<0.105\,\rm{eV}$ at 95\% CL, while the bound on the axion mass is improved to $m_a<1.04\,\rm{eV}$ at 95\% CL.  In this axion-pion thermalization case, the most constraining upper limit on the total neutrino mass lies extremely close to the inverted neutrino mass ordering prediction, enforcing our main message, that is, a multi-messenger search of axions and neutrino properties and for a joint analysis of their expected sensitivities.

Figure~\ref{fig:figure3} illustrates our cosmological constraints in the axion mass - axion-photon coupling plane $(m_a$, $g_{a\gamma})$. We focus exclusively on the parameter space of interest for thermal axions\footnote{For a review of the limits on axion-like particles covering larger ranges see ~\citet{ciaran_o_hare_2020_3932430} and also~\citet{2020arXiv201003889O,OHare:2020wum,Dafni:2018tvj,Knirck:2018knd} for interesting discussions.}, covering a mass range $m_a\in[10^{-6}\,,\,10]\,\rm{eV}$ and quoting our most constraining 95\%~CL bounds for the two thermalization channels together with current experimental limits and future detection sensitivity forecasts. From the limits depicted in Fig.~\ref{fig:figure3} one can notice that a significant range of the parameter space can be probed by cosmological data. Furthermore, a future cosmology-independent limit on the axion mass may provide an important test of the cosmological constraint, and also can be translated into a limit on the hot dark matter component in the form of massive neutrinos, strongly supporting multi-messenger searches of axions and neutrino properties.

\section{Conclusions}\label{sec.Conclusions} 
Axion still provide the most compelling solution to the strong CP problem. Such elusive particles, if realized in nature, can be thermally produced in the primordial Universe, providing a contribution to the hot dark matter component. In light of the last release of CMB temperature and polarization data by the Planck Collaboration, in this paper we improve the existing bounds on thermal relic particles considering two distinct thermal production processes. 

We first focus on the thermal axion population generated by interactions with gluons. Analyzing an extension of the standard $\Lambda$CDM model which includes both axions and massive neutrinos as hot thermal relics and exploiting a combination of Planck CMB temperature, polarization and lensing data, Baryon Acoustic Oscillations measurements, type Ia Supernovae (SNeIa) distance moduli measurements from the Pantheon sample, Galaxy clustering and cosmic shear measurements from the Dark Energy Survey and the prior on the Hubble constant as measured by the SH0ES collaboration from the Hubble Space Telescope data, we derive tight bounds on the thermal hot relic masses. We constrain the axion mass range to $m_a\lesssim 8$ eV at 95\% CL, while the most constraining bound on the total neutrino mass obtained from our numerical analyses is $\sum m_{\nu} \lesssim 0.11 \, \rm{eV}$ at 95\% CL. Interestingly, it lies very close to the minimum value within the inverted mass ordering implied by neutrino oscillation results. 

Then, we perform a model-independent analysis of the axion-pion thermalization channel. Using the same cosmological datasets and restricting our analysis to the parameter space where the chiral Effective Field Theory approach works appropriately, we analyze an extension of the standard $\Lambda$CDM model which includes both axions and massive neutrinos. Without assuming any specific model for the axion-pion interactions, we derive robust constraints on hot relic masses with our most constraining bounds resulting in $m_a\lesssim 0.9\, \rm{eV}$ and $\sum m_{\nu} \lesssim 0.1 \, \rm{eV}$, both at at 95\% CL. Once again the bound on the total neutrino mass is extremely close to the minimum value expected by the inverted mass ordering implied by neutrino oscillation results.

We conclude that scenarios with thermal axions with $m_a\gtrsim 1$~eV would favor the normal ordering as the one governing the mass pattern of neutral fermions, provided axions are thermalized via axion-pion scatterings in the early universe, due to the strong degeneracy between the thermal axion and the neutrino masses. Also future, cosmology-independent probes of neutrino masses may have a huge impact on axion searches. For instance, if the neutrino mass ordering turns out be the inverted one, a wide region in the currently allowed sub-eV thermal axion window would be ruled out. Cosmological and terrestrial laboratory searches for hot thermal relics are therefore complementary and absolutely both required to weigh these particles.

\section*{Acknowledgements}
W.G. and A.M. are supported by "Theoretical Astroparticle Physics" (TAsP), iniziativa specifica INFN. E.D.V. acknowledges the support of the Addison-Wheeler Fellowship awarded by the Institute of Advanced Study at Durham University. O.M. is supported by the Spanish grants FPA 2017-85985-P, PROMETEO/2019/083 and by the European ITN project HIDDeN (H2020-MSCA-ITN-2019//860881-HIDDeN). \\
We thank Luca Visinelli for the interesting suggestions.
In this work we made use of the following \texttt{python} packages that are not mentioned in the text : \texttt{SciPy} \citep{2020SciPy-NMeth} for numerical sampling of the statistical distributions involved in our data analysis, \texttt{GetDist} \citep{2019arXiv191013970L} a tool for the analysis of MCMC samples which employs \texttt{Matplotlib} \citep{Matplotlib} for the realization of the plots in the paper and \texttt{NumPy} \citep{NumPy} for numerical linear algebra.
\section*{Data Availability}
All the data underlying this article are publicly available.


\bibliographystyle{mnras}
\bibliography{main} 




\bsp	
\label{lastpage}
\end{document}